\documentclass[aps,prd,reprint,superscriptaddress]{revtex4-1}
\usepackage{graphicx}
\usepackage{amssymb}
\usepackage{amsmath}
\usepackage{bm}
\usepackage{verbatim}

\usepackage{enumitem} 

\begin{document}

\title{Forming doublons by a quantum quench}

\author{M. Schecter}
\affiliation{School of Physics and Astronomy, University of Minnesota,
Minneapolis, MN 55455, USA}

\author{A. Kamenev}
\affiliation{School of Physics and Astronomy, University
of Minnesota, Minneapolis, MN 55455, USA}
\affiliation{William I. Fine
Theoretical Physics Institute, University of Minnesota, Minneapolis, MN
55455, USA}

\date{\today}

\begin{abstract}
Repulsive interactions between particles on a lattice may lead to bound states, so called doublons. Such states may be created by dynamically tuning the interaction strength, {\em e.g.} using a Feshbach resonance, from attraction to repulsion. We study the doublon production efficiency as a function of the tuning rate at which the on-site interaction is varied. An expectation based on the Landau-Zener law suggests that exponentially few doublons
are created in the adiabatic limit. Contrary to such an expectation, we found that the number of produced doublons scales as a power law of the tuning rate with the exponent dependent on the dimensionality of the lattice. The physical reason for this anomaly is the effective decoupling of doublons from the two-particle continuum for center of mass momenta close to the corners of the Brillouin zone. The study of doublon production may be 
a sensitive tool to extract detailed information about the band structure.  
\end{abstract}
\maketitle

\emph{Introduction} --  Quantum evolution of a system driven far from equilibrium by time-dependent perturbations 
became a focus of increased attention with the advent of cold atomic gases.
These systems provide a unique advantage due to their high degree of tunability vis-a-vis various system parameters including dimensionality, density, inter-particle interaction strengths and disorder \cite{Bloch_Zwerger_Cold_Atoms_2008}. In particular, such systems provide remarkably accurate realizations of both Bose and Fermi Hubbard models, in which the rich equilibrium phase diagrams \cite{Jaksch_Hubbard_phases_1998} or far-from-equilibrium dynamics may be studied in a highly controllable manner.

For strong attraction, spin$-1/2$ fermions preferentially pair on lattice sites forming tightly bound composite bosons. A slow quench of the interaction strength takes the system through the BEC-BCS crossover into a regime of strong repulsion, resulting in a ground state which is a Fermi liquid or Mott insulator, depending on the filling fraction \cite{Esslinger_Fermi_Hubbard}. However, if such a quench is sufficiently fast, a fraction of initial pairs may not have time to dissociate and thus transform into repulsively bound states $-$ {\em doublons} \cite{Winkler_Doublon_Nature_2006}. Although being high-energy states, doublons have a long lifetime due to energy conservation, which requires a coherent multi-particle excitation to induce doublon decay. The lifetime of residual doublons was investigated for  both  bosonic and fermionic counterparts  \cite{Esslinger_Demler_Doublon_PRL_2010,Kamenev_Doublon_decay_2011}.  
Here we focus on the creation efficiency of such repulsively-bounded pairs \cite{Esslinger_Demler_Doublon_PRL_2010,Kamenev_Doublon_decay_2011,Winkler_Doublon_Nature_2006}, upon a quench of the on-site interaction strength. The latter may be achieved, using \emph{e.g.} a Feshbach resonance \cite{Kohler_Feshbach_2006}, and is characterized by a rate $\lambda$ dictating the speed of the ramp for the on-site particle interaction $U=\lambda t$. After the quench a fraction of remaining double occupancies, or production efficiency of doublons, depends sensitively on the ramp rate: for fast ramps most doublons survive, while in the adiabatic limit the majority softly dissociate into free particles  states forming a Bloch band.

\begin{figure}[t]
\includegraphics[width=\columnwidth]{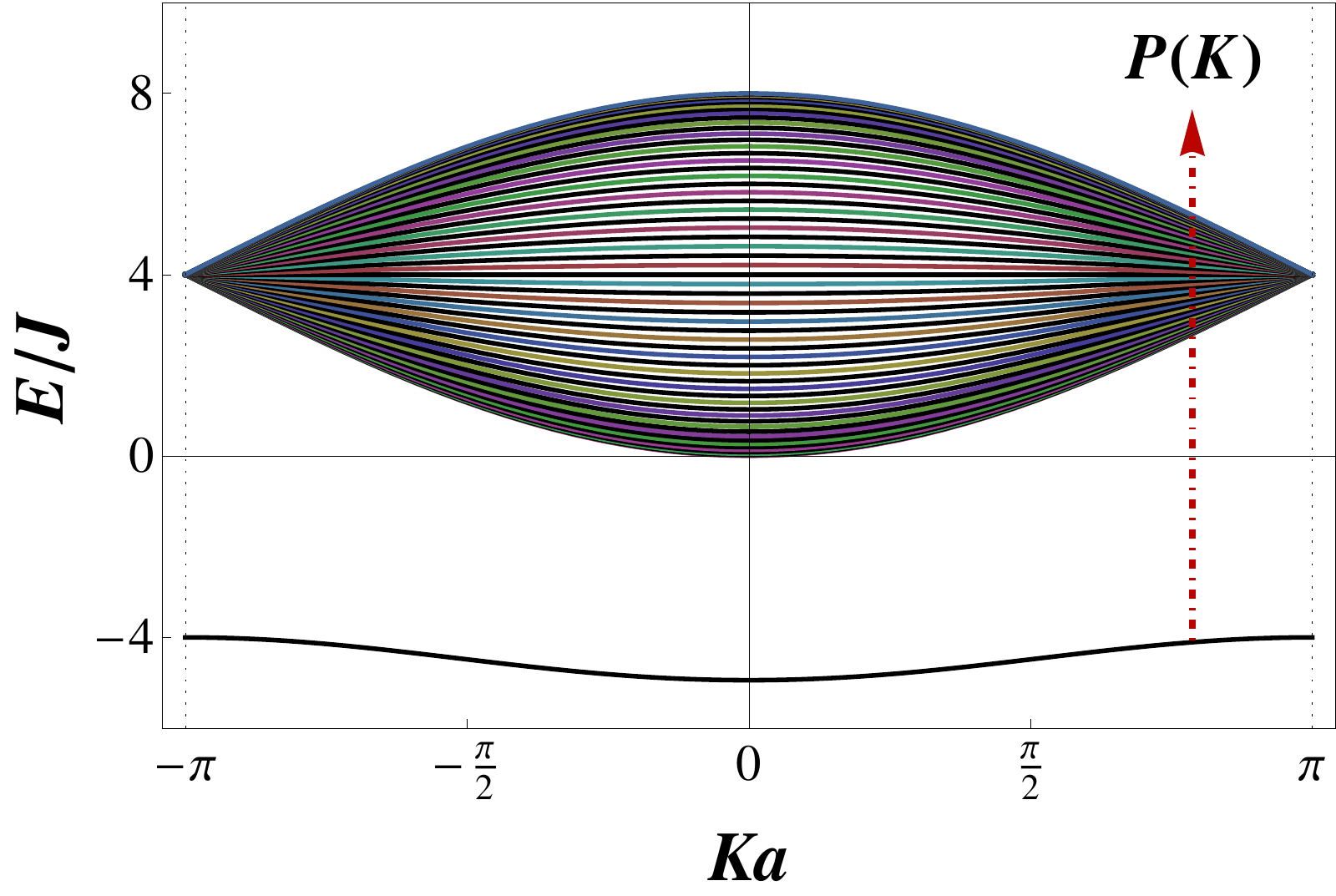}
\caption{(Color online) Two-particle spectrum and doublon bound state energy as a function of the center of mass quasi-momentum $K$ for negative on-site interaction $U$. Due to quasi-translational invariance, doublons evolve along curves of constant $K$ and survive the ramp with probability $\mathcal{P}(K)$. Near the corners of the Brillouin zone, $|K|a=\pi$, the two-particle bandwidth and doublon-band coupling tend to zero, rendering those states most likely to surive the ramp.}
\label{fig:spectrum}
\end{figure}

Our main results for the doublon production efficiency are as follows. In the limit of small initial concentration of attractively bound pairs the problem may be treated in the spirit of a Landau-Zener crossing \cite{shytov,sinitsyn,Brundobler_Elser} of a time-dependent level and a time-independent band.  Such a variant of the Landau-Zener problem is known as the Demkov--Osherov model \cite{DO,Brundobler_Elser}, which admits an exact solution. In terms of the rate $\lambda$ and nearest-neighbor hopping $J_i$ along the primitive vectors of a $D-$dimensional lattice, we found for the probability for the two particles to remain on the same site
after a quench ({\em e.g.} for $t\to +\infty$)
\begin{eqnarray}
\label{eq:prob}
\nonumber\mathcal{P}&=&\prod_{i=1}^{D}I_{0}(8\pi J_{i}^{2}/\lambda)e^{-8\pi J_{i}^{2}/\lambda},\\&=&\begin{cases}
1-\frac{8\pi}{\lambda}\sum_{i}J_{i}^{2}; & \lambda\gg8\pi J_i^2\\
\left(\frac{\lambda}{16\pi^2 J^2}\right)^{D/2}; & \lambda\ll8\pi J^2,\end{cases}
\end{eqnarray}
where $J=\left(\prod_i J_i\right)^{\frac{1}{D}}$ is the geometric mean hopping, $I_0$ is a modified Bessel function of the first kind and the asymptotic limits are provided in the second line. Remarkably, the doublon survival probability is \emph{not} exponentially suppressed in the adiabatic limit, like $\mathcal{P}\sim \prod_i e^{-8\pi J_i^2/\lambda}$, but rather scales as a power law $\mathcal{P}\sim(\lambda/J^2)^{D/2}$. This implies that doublons have a significantly better chance of surviving the quench than might be naively expected. As explained below, the power-law survival enhancement comes from the presence of doublons with center of mass quasi-momenta near the Brillioun zone corners, see Fig.~\ref{fig:spectrum}. We also considered the case of a finite concentration of initially attractive bounded pairs, taking into consideration the bosonic or fermionic nature of dissociated particles. We show that particle statistics renormalizes the production efficiency in an intuitive way: for fermions, Pauli blocking of dissociation channels leads to doublon production enhancement, while for bosons stimulated dissociation suppresses doublon production.

\emph{Dilute limit} -- In the presence of a sufficiently deep optical lattice one may neglect higher Bloch bands and focus entirely on the lowest. This approximation leads to the Hubbard model characterized by hopping $J_i$ along the primitive vectors and on-site interaction energy $U$. Below we express the corresponding Hamiltonian assuming particles are spin$-\frac{1}{2}$ fermions. The spinless bosonic counterpart is obtained by removing spin indices and corresponding summations.
\begin{equation}
\label{eq:hamil}
H=-\sum_{\langle i,j\rangle\sigma}J_{ij}c_{i\sigma}^{\dagger}c_{j\sigma}+U(t)\sum_{i}n_{i\uparrow}n_{i,\downarrow}-\mu\sum_{i}n_{i}
\end{equation}
where $\langle i,j\rangle$ restricts summation to nearest neighbor sites, $\sigma=\pm1$ denotes the spin, $c^{\dagger}\,(c)$ are fermionic creation (annhilation) operators and $n_{j\sigma}=c^\dagger_{j\sigma} c_{j\sigma}$ is the number operator. Below we focus on a rectangular lattice with lattice constants $a_i$ and hopping $J_i$ ($i=1,2,3$).

\begin{figure}[t]
\includegraphics[width=\columnwidth]{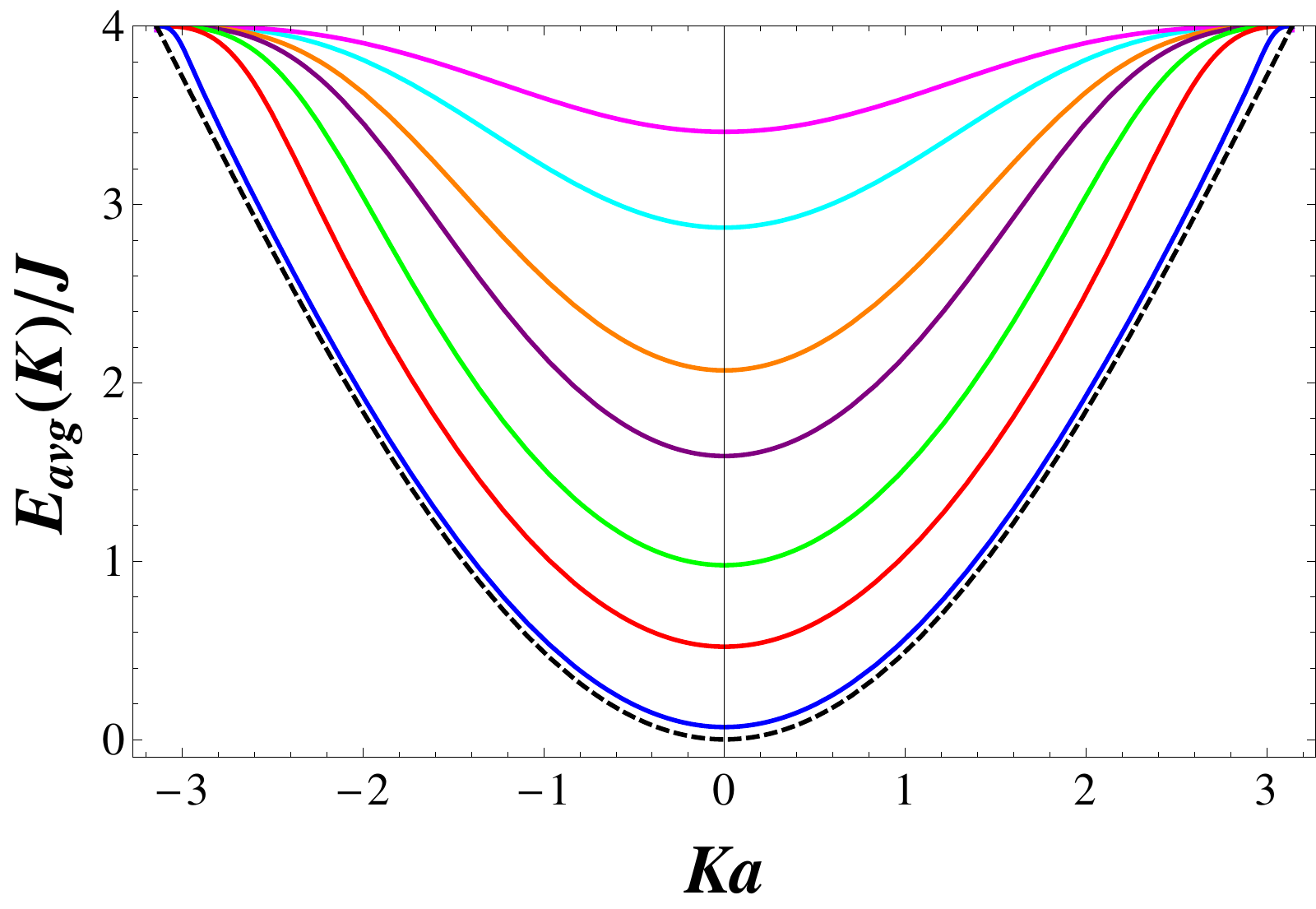}
\includegraphics[width=\columnwidth]{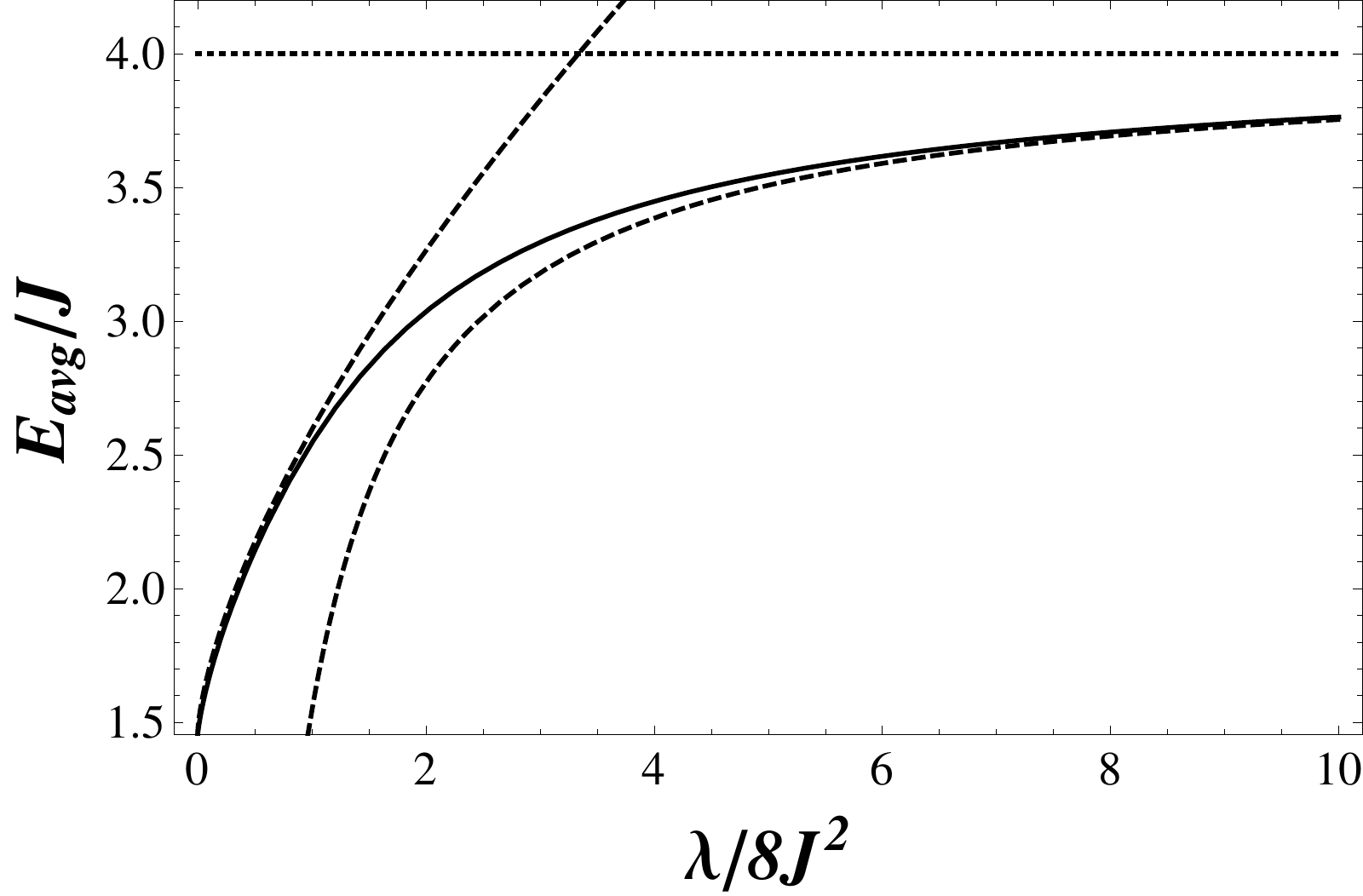}
\caption{(Color online) Top panel: average two-particle energy of a dissociated doublon as a function of the center of mass quasi-momentum $K$. Each curve is plotted with a different value of the parameter $\lambda/8J^2$ given by (from top to bottom): 6, 3, 1.5, 1, 0.5, 0.2, 0.01. The dashed curve is the lower edge of the two-particle continuum. Bottom panel: average dissociation energy integrated over $K$ as a function of the rate. Dashed lines correspond to the asymptotic limits given by Eq.~(\ref{eq:energy_avg}).}
\label{fig:energy_avg}
\end{figure}

In the case of two particles one may write a wavefunction of their coordinates $\Psi(\textbf{x}_1,\textbf{x}_2)$ in terms of central and relative coordinates as $\Psi(\textbf{x}_1,\textbf{x}_2)=e^{i\textbf{K}\textbf{R}}\psi_{\textbf{K}}(\textbf{r})$, where $\textbf{R}=(\textbf{x}_1+\textbf{x}_2)/2$, $\textbf{r}=\textbf{x}_1-\textbf{x}_2$ and $\textbf{K}$ is the center of mass quasi-momentum of the particle pair. In the case of bosons, one requires the relative wavefunction $\psi_{\mathbf{K}}(\mathbf{r})$ to be an even function of the relative separation $\mathbf{r}$. The same is also true of spin$-\frac{1}{2}$ fermions occupying the spin-singlet state (on the triplet manifold $\psi_{\mathbf{K}}(\mathbf{r})$ is odd and thus on-site interaction is irrelevant). Consequently, the two-particle results derived below extend to either species of constituent particles. We shall find, unsurprisingly, that particle statistics do play a role in the many-body situation considered later.

Due to translation invariance of lattice systems, we work in a sector with fixed $\textbf{K}$. The action of Hamiltonian (\ref{eq:hamil}) on the two-particle wavefunction $e^{i\textbf{K}\textbf{R}}\psi_{\textbf{K}}(\textbf{r})$ produces an effective tight-binding model in the relative coordinate $\textbf{r}$
\begin{equation}
\label{eq:hamil_r}
\left(-\bar{\textbf{J}}\cdot\Delta_{\textbf{r}}+E_{\mathbf{K}}+U\delta_{\textbf{r},0}\right)\psi_{\textbf{K}}(\textbf{r})=E\psi_{\textbf{K}}(\textbf{r}),
\end{equation}
where we introduced an effective hopping with components $\bar{\textbf{J}}_i=2J_i\mathrm{cos}\frac{\textbf{K}_i a_i}{2}$, the discrete Laplacian $\left[\Delta_{\textbf{r}}\psi_{\textbf{K}}(\textbf{r})\right]_i=\psi_{\textbf{K}}(\textbf{r}+\textbf{a}_i)+\psi_{\textbf{K}}(\textbf{r}-\textbf{a}_i)-2\psi_{\mathbf{K}}(\mathbf{r})$ and the center of mass kinetic energy $E_{\mathbf{K}}=\sum_i 4J_i(1-\mathrm{cos}\frac{\mathbf{K}_i\mathbf{a}_i}{2})$.

An interesting feature of Eq.~(\ref{eq:hamil_r}) is that it supports a bound state solution at both negative and \emph{positive} values of the on-site interaction $U$. This can be seen most easily introducing the Fourier transformation $\psi_{\textbf{K}}(\textbf{r})=\sum_{\textbf{q}} e^{i\textbf{q}\textbf{r}}\psi_{\textbf{K}}(\textbf{q})$ into Eq.~(\ref{eq:hamil_r}), which gives rise to an equation for the bound state energy $E_\mathrm{d}$
\begin{equation}
\label{eq:energy}
1=\int\frac{d^D\textbf{q}}{(2\pi)^D}\frac{U}{E_{\mathrm{d}}-\varepsilon(\textbf{K},\textbf{q})}.
\end{equation}
Here $\varepsilon(\textbf{K},\textbf{q})=\sum_i 4J_i\left[1-\mathrm{cos}\frac{\textbf{K}_ia_i}{2}\mathrm{cos}\left(\textbf{q}_ia_i\right)\right]$ is the two-particle energy spectrum, while Eq.~(\ref{eq:energy}) may be viewed as a condition of the dressed Green function having a pole on the real axis at $E=E_{\mathrm{d}}$ \cite{Winkler_Doublon_Nature_2006,Molmer}. In one spatial dimension (\emph{e.g.,} $\bar{\textbf{J}}_x=\bar{\textbf{J}}_y=0$), the particular form of the two-particle spectrum allows to integrate Eq.~(\ref{eq:energy}) explicitly, giving the bound state energy $E_{\mathrm{d}}(K)=4J+\mathrm{sgn}(U)\sqrt{U^{2}+16J^{2}\mathrm{cos}^{2}\frac{Ka}{2}}$, seen in Fig.~\ref{fig:spectrum}. The existence of a bound state for \emph{any} value of $U\neq0$ extends also to $D=2$, where the density of states $\mathcal{N}(\varepsilon)$ near the band bottom is constant $\mathcal{N}(\varepsilon)\to\mathcal{N}_0$. Similar to the Cooper problem, the bound state energy $E_{\mathrm{d}}\sim-Je^{-1/\mathcal{N}_0U}$ is thus exponentially shallow as $U\to0$. In $D=3$ the density of states vanishes at the band bottom, resulting in the convergence of the momentum integral at $E_{\mathrm{d}}=0$ and thus leads to a critical value $U_{\mathrm{crit}}(\mathbf{K})=\left|\int \frac{d^3\mathbf{q}}{2\pi^3}\frac{1}{\varepsilon(\mathbf{K},\mathbf{q})}\right|^{-1}$ such that for $|U|<U_{\mathrm{crit}}(\mathbf{K})$ no bound state exists for a given $\mathbf{K}$.

Another intriguing property of Eq.~(\ref{eq:hamil_r}) is the dependence of the effective hopping on the center of mass momentum: $\bar{\textbf{J}}_i=2J_i\textrm{cos}\left(\textbf{K}_ia_i/2\right)$. Consequently, the hopping along a particular direction is identically \emph{zero} on the corresponding Brillouin zone surface defined by $|\textbf{K}_i|=\pi/a_i$. This is the result of destructive interference between two amplitudes that both increase the relative coordinate $\textbf{r}$ along the specified direction, with one shifting the central coordinate by $+a_i/2$, the other by $-a_i/2$. The presence of zero hopping is also manifest in the two-particle band spectrum (see Fig.~\ref{fig:spectrum} for $D=1$) which becomes pinched at the BZ boundaries, leading to a system of decoupled (and hence degenerate) states. Such momentum dependent hopping was discussed for $D=1$ in Refs.~\cite{Molmer,Stecher} in the context of Feshbach molecules coupled to a band continuum.

We proceed by transforming Eq.~(\ref{eq:hamil_r}) into a form allowing the results of the Demkov--Osherov (DO) model to be conveniently applied. This is achieved by introducing a unitary transformation $\psi_{\textbf{K}}(\textbf{r})=\psi_{\mathbf{K}}(\mathbf{r})\delta_{\mathbf{r,0}}+\sum_{\mathbf{q}}\mathcal{U}_{\textbf{r},\mathbf{q}}\tilde{\psi}_{\textbf{K},\mathbf{q}}$ which diagonalizes Eq.~(\ref{eq:hamil_r}) in the subspace of states having $\textbf{r}\neq\mathbf{0}$ (\emph{i.e.,} the unitary transformation satisfies boundary condition $\mathcal{U}_{\mathbf{0},\mathbf{q}}=\mathcal{U}_{\mathbf{q},\mathbf{0}}=0$),
\begin{equation}
\label{eq:hamil_diag}
\left(\begin{array}{cc}
\lambda t & \tilde{J}_{\mathbf{K},\mathbf{q}}\\
\tilde{J}_{\mathbf{K},\mathbf{q}}^{\dagger} & \varepsilon_{\mathbf{K},\mathbf{q}}\end{array}\right)\left(\begin{array}{c}
\psi_{\mathbf{K}}(\mathbf{0})\\
\tilde\psi_{\mathbf{K},\mathbf{q}}\end{array}\right)=E\left(\begin{array}{c}
\psi_{\mathbf{K}}(\mathbf{0})\\
\tilde\psi_{\mathbf{K},\mathbf{q}}\end{array}\right)
\end{equation}
where $\tilde{J}_{\mathbf{K},\mathbf{q}}=-2\sum_i J_i\mathrm{cos}\left(\textbf{K}_ia_i/2\right)\mathcal{U}_{\mathbf{a}_i,\mathbf{q}}$ and $\varepsilon_{\mathbf{K},\mathbf{q}}$ is the corresponding set of band energies. Since we deal with a single site defect in an infinite system, the spectrum $\varepsilon_{\mathbf{K},\mathbf{q}}$ coincides with the two-particle spectrum $\varepsilon(\mathbf{K},\mathbf{q})$ given above. Indeed, the most important aspects of the site defect, its energy $U=\lambda t$ and coupling to band states, has been included explicitly. We note that, owing to the point boundary condition, the corresponding eigenfunctions $\tilde{\psi}_{\mathbf{K},\mathbf{q}}$ are not simply plane waves except for the special case $D=1$ (see below).

The Hamiltonian (\ref{eq:hamil_diag}), being now of the DO form, allows to read off the exact survival probability of the $\mathbf{r}=\mathbf{0}$ state: $\mathcal{P}(\mathbf{K})=\prod_{\mathbf{q}}e^{-2\pi\left|\tilde{J}_{\mathbf{K},\mathbf{q}}\right|^2/\lambda}$ \cite{DO,Brundobler_Elser}. Without a precise knowledge of the unitary matrix $\mathcal{U}$, such a result is seemingly quite useless. Remarkably, however, one can now determine the survival probability explicitly \emph{without} knowing the precise form of the unitary transformation. Raising the product over states to summation in the exponent, one may show using only the unitarity property $\mathcal{U}^\dagger\mathcal{U}=\mathbf{\hat{1}}$ that
\begin{equation}
\label{eq:prob_K}
\mathcal{P}(\mathbf{K})=\prod_{i}\mathrm{exp}\left[-\frac{16\pi\left|J_i\right|^2}{\lambda}\mathrm{cos}^2\left(\frac{\mathbf{K}_ia_i}{2}\right)\right].
\end{equation}
Within the above LZ framework the initial ($t=-\infty$) on-site energy is taken to be $U/J=-\infty$. This implies that initially doublons, having infinite effective mass, are in \emph{site} eigenstates. The initial state is thus one whose momenta $\mathbf{K}$ components are uniformly distributed over the entire Brillouin zone \cite{footnote1}. As a result, the doublon survival probability is given by the (evenly weighted) average over the BZ, namely $\mathcal{P}=\sum_{\mathbf{K}}\mathcal{P}(\mathbf{K})$. Summing Eq.~(\ref{eq:prob_K}) over $\mathbf{K}$ gives the main result Eq.~(\ref{eq:prob}) \cite{footnote}.

After the ramp there exists, in addition to a fraction of surviving doublons, a fraction of dissociated, or non-surviving, particles occupying band states. One may ask what is the non-equilibrium distribution of such particles. To this end we require the probability of doublon dissociation into state $\mathbf{q}$: $\mathcal{P}_{\mathbf{q}}(\mathbf{K})=(1-z_{\mathbf{q}})\prod_{\varepsilon_{\mathbf{k}}<\varepsilon_{\mathbf{q}}}z_{\mathbf{k}}$, where $z_{\mathbf{q}}=e^{-2\pi |\tilde{J}_{\mathbf{K},\mathbf{q}}|^2/\lambda}$ \cite{DO,Brundobler_Elser}. This distribution allows to study, {\em e.g.} the average energy of non-surviving particles in the band. For fixed $\mathbf{K}$ the latter is written as $E_{\mathrm{avg}}(\mathbf{K})=\sum_{\mathbf{q}} \mathcal{P}_{\mathbf{q}}(\mathbf{K})\varepsilon(\mathbf{K},\mathbf{q})/(1-\mathcal{P}(\mathbf{K}))$. In the adiabatic limit, $\lambda\to0$, $z_{\mathbf{q}}$ is exponentially small, and as a result of the energy summation above, we find $\mathcal{P}_{\mathbf{q}}(\mathbf{K})\to\delta_{\mathbf{q},\mathbf{0}}$, where $\varepsilon(\mathbf{K},\mathbf{0})$ is the lower edge of the two-particle continuum, Fig.~\ref{fig:spectrum}. In the same limit $\mathcal{P}(\mathbf{K})\to0$ so that $E_{\mathrm{avg}}(\mathbf{K})\to\sum_i 4J_i(1-\mathrm{cos}\frac{\mathbf{K}_i a_i}{2})=E_{\mathbf{K}}$ (\emph{i.e.,} in the adiabatic limit the two-particle energy is simply the center of mass kinetic energy of the dissociated doublon). In the diabatic limit, $\lambda\to\infty$, one may use $\mathcal{P}_{\mathbf{q}}(\mathbf{K})=2\pi|\tilde{J}_{\mathbf{K},\mathbf{q}}|^2/\lambda$ along with $H_{\mathbf{a}_i,\mathbf{a}_j}=\left[\mathcal{U}\mathrm{diag}\{\varepsilon_{\mathbf{K},\mathbf{q}}\}\mathcal{U}^\dagger\right]_{\mathbf{a}_i,\mathbf{a}_j}=\delta_{\mathbf{a}_i,\mathbf{a}_j}\sum_k 4J_k $ to calculate the average dissociation energy, $E_{\mathrm{avg}}(\mathbf{K})=\sum_i 4J_i$, which is nothing but the band center energy. To obtain results between the two asymptotic limits above, one must have access to the explicit form of the unitary matrix $\mathcal{U}_{\mathbf{a}_i,\mathbf{q}}$.

For simplicity of calculation, we solved for the unitary transformation  in $D=1$, where the point boundary condition coincides with the surface boundary condition familiar from a particle in a box. Hence, $\mathcal{U}_{r,q}=\sqrt{\frac{2}{N}}\mathrm{sin}(rq)$ where $q$ may be identified with the relative momentum of band particles, since the eigenfunctions are odd combinations of plane waves. Taking the continuum limit $N\to\infty$, one finds the probability of dissociating into the range $(q,q+dq)$,
\begin{eqnarray}
\label{eq:mom_dist}
\nonumber\mathcal{P}_{q}(K)&=&\frac{16J^{2}}{\lambda}\mathrm{cos}^{2}\left(\frac{Ka}{2}\right)\mathrm{sin}^{2}(qa)\\&\times& \mathrm{Exp}\left[-\frac{16J^{2}}{\lambda}\mathrm{cos}^{2}\frac{Ka}{2}\left|qa-\mathrm{sin}qa\,\mathrm{cos}qa\right|\right].
\end{eqnarray}
One notices that $\mathcal{P}_q(K)|_{q=0,\pm\pi}=0$, implying that doublons are effectively decoupled from states with these relative momenta. This is unsurprising considering the relative group velocity $v_q=\partial_q\varepsilon(K,q)$ vanishes at these points, \emph{i.e.,} if the relative velocity is zero there can be no dissociation. Using Eq.~(\ref{eq:mom_dist}) one may now calculate $E_{\mathrm{avg}}(K)$ for various $\lambda$, shown in Fig.~\ref{fig:energy_avg}.

To monitor ramp-induced heating, one may study the  \emph{total} dissociation energy $E_{\mathrm{avg}}=\sum_{\mathbf{K}} E_{\mathrm{avg}}(\mathbf{K})$. For arbitrary $D$ the asymptotics of this quantity lie in the range dictated by the $\lambda$-independent limits of $E_{\mathrm{avg}}(\mathbf{K})$ given above. For $D=1$ we determine the leading dependence on $\lambda$ in the vicinity of these bounds shown in Fig.~\ref{fig:energy_avg},
\begin{equation}
\label{eq:energy_avg}
E_{\mathrm{avg}}=4J-\frac{8J}{\pi}\begin{cases}
C\left(\frac{\lambda}{8J^{2}}\right)^{-1}; & \lambda\gg8J^{2}\\
1-C^\prime\left(\frac{\lambda}{8J^{2}}\right)^{2/3}; & \lambda\ll8J^{2},\end{cases}
\end{equation}
with numerical factors $C=0.966$ and $C^\prime=0.448$. Equation~(\ref{eq:energy_avg}) implies that the system is heated by an amount of order $J$ \emph{regardless} of how slow the ramp is.

\emph{Finite doublon concentration} -- To go beyond the above two-particle physics we study the secondary quantized Hamiltonian $H=H_0+H_{\mathrm{int}}$ describing the dynamics of doublons coupled to particles occupying band states (for simplicity we hereafter set the band center to be the zero of energy). Below we express equations for spin$-\frac{1}{2}$ fermions, but provide results also for spinless bosons.
\begin{eqnarray}
\label{eq:doublon_ham}
H_{0}&=&U(t)\sum_{\mathbf{K}}d_{\mathbf{K}}^{\dagger}d_{\mathbf{K}}-\sum_{\mathbf{q},\sigma,i}2J_{i}\mathrm{cos}(\mathbf{q}_{i}\mathbf{a}_{i})c_{\mathbf{q}\sigma}^{\dagger}c_{\mathbf{q}\sigma},\\H_{\mathrm{int}}&=&\sum_{\mathbf{K},\mathbf{q},\sigma}\sigma\tilde{J}_{\mathbf{K},\mathbf{q}}d_{\mathbf{K}}^{\dagger}c_{\frac{\mathbf{K}}{2}+\mathbf{q},\sigma}c_{\frac{\mathbf{K}}{2}-\mathbf{q},-\sigma}+\mathrm{h.c.}
\end{eqnarray}
The interaction Hamiltonian $H_{\mathrm{int}}$ describes the dissociation and association of (bosonic) doublons with total momentum $\mathbf{K}$ onto two particles with momenta $\frac{\mathbf{K}}{2}\pm \mathbf{q}$. Notice that in the form Eq.~(\ref{eq:doublon_ham}), the doublon band is completely flat $E(\mathbf{K})=U$. To obtain the renormalized doublon energy $E_{\mathrm{d}}$, one integrates out fermionic degrees of freedom and expands the resulting action to second order in $d_{\mathbf{K}}$ operators, $d^\dagger_{\mathbf{K}}(\varepsilon)\left[\varepsilon-U-\sum_{\mathbf{q}}\frac{\left|\tilde{J}_{\mathbf{K},\mathbf{q}}\right|^2}{\varepsilon-\varepsilon(\mathbf{K},\mathbf{q})}\right]d_{\mathbf{K}}(\varepsilon)=d^\dagger_{\mathbf{K}}(\varepsilon)\mathcal{D}^{-1}(\mathbf{K},\varepsilon)d_{\mathbf{K}}(\varepsilon)$. The doublon bound state energy is defined by the pole of the corresponding propagator, $\mathcal{D}^{-1}(\mathbf{K},E_{\mathrm{d}})=0$, which also shares the solution to Eq.~(\ref{eq:energy}).

Following Ref.~\cite{Altland_Gurarie}, we derive a set of coupled quantum kinetic equations governing the distribution functions $n_{\mathrm{d}}=n_{\mathrm{d}}(\mathbf{K},t)$ for doublons and $n_{\mathbf{q}\sigma}=n_{\mathbf{q}\sigma}(t)$ for free particles,
\begin{eqnarray}
\label{eq:kinetic_eq1}
\dot{n}_{\mathrm{d}}&=&-\pi\sum_{\mathbf{q},\sigma}|\tilde{J}_{\mathbf{K},\mathbf{q}}|^{2}\delta(\lambda t-\varepsilon(\mathbf{K},\mathbf{q}))\\\nonumber&\times&\left\{ n_{\mathrm{d}}\left(1-n_{\mathbf{\frac{K}{2}}+\mathbf{q},\sigma}-n_{\frac{\mathbf{K}}{2}-\mathbf{q},-\sigma}\right)-n_{\mathbf{\frac{K}{2}}+\mathbf{q},\sigma}n_{\frac{\mathbf{K}}{2}-\mathbf{q},-\sigma}\right\}
\\\nonumber\dot{n}_{\mathbf{q}\sigma}&=&-2\pi\sum_{\mathbf{K}}|\tilde{J}_{\mathbf{K},\mathbf{q}-\mathbf{K}/2})|^{2}\delta(\lambda t-\varepsilon(\mathbf{K},\mathbf{q}-\mathbf{K}/2))\\\label{eq:kinetic_eq2}&\times&\left\{ n_{\mathbf{q}\sigma}n_{\mathbf{K}-\mathbf{q},-\sigma}-n_{\mathrm{d}}\left(1-n_{\mathbf{q}\sigma}-n_{\mathbf{K}-\mathbf{q},-\sigma}\right)\right\} .
\end{eqnarray}
The factor in curly brackets in Eqs.~(\ref{eq:kinetic_eq1}), (\ref{eq:kinetic_eq2}) comes from the combination of "in" and "out" terms: $n_{\frac{\mathbf{K}}{2}+\mathbf{q}\sigma}n_{\frac{\mathbf{K}}{2}-\mathbf{q},-\sigma}\left(1+n_{\mathrm{d}}\right)-n_{\mathrm{d}}\left(1- n_{\frac{\mathbf{K}}{2}+\mathbf{q}\sigma}\right)\left(1-n_{\frac{\mathbf{K}}{2}-\mathbf{q},-\sigma}\right)$. The latter piece corresponds to the "out" process, where $1-n_{\frac{\mathbf{K}}{2}\pm\mathbf{q},\pm\sigma}$ reflects the fact that doublon decay is prohibited if either free particle state is occupied, \emph{i.e.} $n_{\frac{\mathbf{K}}{2}\pm \mathbf{q},\pm\sigma}=1$. For bosons, the minus signs in front of $n_{\mathbf{q},\sigma}$ are changed to plus, replacing Pauli blocking with stimulated emission. The "in" process may be understood on the same grounds.

In the single doublon, vanishing concentration, limit ($n_{\mathbf{q}\sigma}=0$), Eq.~(\ref{eq:kinetic_eq1}) indeed posseses the LZ solution: $n_{\mathrm{d}}(\mathbf{K},+\infty)/n_{\mathrm{d}}(\mathbf{K},-\infty)=e^{-2\pi\sum_{\mathbf{q}}|\tilde{J}_{\mathbf{K},\mathbf{q}}|^2/\lambda}$, Eq.~(\ref{eq:prob_K}). The aim of the above outlined kinetic scheme is thus to obtain leading order corrections to the efficiency in the presence of finite doublon concentration. In the rapid ramp limit, $J^2\ll \lambda$, we solve the coupled kinetic equations by first substituting the time-independent initial doublon distribution $n_{\mathrm{d}}=n_{\mathrm{d}}^{(i)}$ into Eq.~(\ref{eq:kinetic_eq2}) and solve for $n_{\mathbf{q}\sigma}$ neglecting non-linear contributions, $n_{\mathbf{q}\sigma}(t)=2\pi n_{\mathrm{d}}^{(i)}\lambda^{-1} \sum_{\mathbf{K}}|\tilde{J}_{\mathbf{K},\mathbf{q}-\mathbf{K}/2}|^2 \Theta(\lambda t-\varepsilon(\mathbf{K},\mathbf{q}-\mathbf{K}/2))$. This approximation amounts to using the conservation law $2\sum_{\mathbf{K}} n_{\mathrm{d}}^{(i)}=2\sum_{\mathbf{K}} n_{\mathrm{d}}^{(f)}+\sum_{\mathbf{q},\sigma} n_{\mathbf{q}\sigma}^{(f)}$ where $n_{\mathrm{d}}^{(f)}=n_{\mathrm{d}}^{(i)}\left(1-2\pi\lambda^{-1}\sum_{\mathbf{q}}|\tilde{J}_{\mathbf{K},\mathbf{q}}|^2\right)$ is calculated in the leading (zero density) order. Substituting the above approximation for $n_{\mathbf{q}\sigma}$ into Eq.~(\ref{eq:kinetic_eq1}) using the $D=1$ unitary matrix $\mathcal{U}$ above gives the final fraction of surviving doublons
\begin{equation}
\label{eq:doublon_fast}
\mathcal{P}=1-\frac{8\pi J^2}{\lambda}+\frac{3}{4}\left(1\pm\frac{3}{4}\nu\right)\left(\frac{8\pi J^2}{\lambda}\right)^2,
\end{equation}
valid to the leading order in $\nu\ll1/2$, where $\pm$ refers to fermionic (bosonic) particles and $\nu$ is the filling fraction related to the initial doublon distribution as $n_{\mathrm{d}}^{(i)}(K)=\nu$. That the leading correction in $\nu$ occurs in second order follows from the fact that one order is required to first populate the free-particle levels and another is needed to act back on the doublon decay rate.

The more interesting (power-law) part of the survival probability occurs in the adiabatic regime. As explained in Ref.~\cite{Altland_Gurarie}, this regime is beyond the scope of the kinetic equation approach. In brief, this is because Eqs.~(\ref{eq:kinetic_eq1}), (\ref{eq:kinetic_eq2}) do not predict evolution from ground state to ground state, namely that all doublons must be exhausted in the formation of a Fermi sea as $U\to+\infty$ and $\lambda\to0$. To properly take this fact into account, we propose the following combined strategy. In the extreme adiabatic limit, only those doublons with momenta components sufficiently close to $|\mathbf{K}_i|=\pi/a_i$ have an appreciable chance of surviving, \emph{i.e.,} an effective coupling small compared to $\sqrt{\lambda}$. This fact allows to treat their dynamics in the diabatic sense, subject to the kinetic theory outlined above. All other doublons with momentum away from the BZ boundary dissociate into states near $q=0$ due to the exponential suppression of their survival probability. From the perspective of doublons near the BZ corners, they observe an essentially uniform distribution of occupied band states $n_{\mathbf{q}\sigma}=\nu$. According to Eq.~(\ref{eq:kinetic_eq1}), this modifies the effective rate as $\lambda\to\lambda/(1\mp2\nu)$. As a result, the power-law of Eq.~(\ref{eq:prob}) becomes
\begin{equation}
\label{eq:prob_bosons}
\mathcal{P}=\left[\frac{\lambda}{\left(1\mp2\nu\right)16\pi^2J^2}\right]^{D/2},
\end{equation}
where $\mp$ refers to fermionic (bosonic) constituent particles.  From Eq.~(\ref{eq:prob_bosons}), we see that the effect of stimulated emission for bosons \emph{decreases} the total survival probability, making the process appear effectively slower, while the effect of Pauli blocking in the case fermions \emph{increases} the survival probability because there is less phase space for doublon dissociation.

\emph{Conclusions} -- The presence of a periodic potential leads to the formation of interband gaps in which a high energy bound state (doublon) may exist. The formation of such doublons may be achieved employing the concept of Feshbach resonance in which the inter-particle coupling strength is changed by applying a time-varying magnetic field. We showed that doublon creation is not exponentially suppressed in the adiabatic limit but rather scales as a power law $\mathcal{P}\sim(\lambda/J^2)^{D/2}$, contrary to conventional LZ wisdom. Beyond the two-particle picture, we derived finite concentration corrections to the survival probability depending on the bosonic or fermionic nature of the constituent particles. Going beyond the dilute limit towards filling factors of order 1/2 requires a more careful treatment of the Hubbard Hamiltonian Eq.~(\ref{eq:hamil}) than provided here. In particular, the bosonic nature of the doublon fields $d_{\mathbf{K}},\,d^\dagger_{\mathbf{K}}$ in Eq.~(\ref{eq:doublon_ham}) can only be justified in the dilute limit where doublons are sparse, their mutual scattering is irrelevant and hence their internal structure (composite nature) is of no importance.

We are grateful to E. Demler and D. Gangardt for stimulating discussions. This work was supported by DOE grant DE-FG02-08ER46482.


\begin{thebibliography}{26}
\bibitem{footnote1} In practice the initial on-site attraction is not infinitely strong and doublons form a thermal distribution in a band with finite bandwidth. However, in typical experiments the temperature greatly exceeds the doublon bandwidth, implying that the distribution is uniform. As a result, the evenly weighted average over the BZ, and hence power-law survival probability, remains in tact.
\bibitem{footnote} In the presence of next-nearest hopping $J^{(2)}_i\ll J_i$, exponential dependence is recovered due to the widening of two-particle spectrum at $|\mathbf{K}|=\pi/a_i$, and hence non-vanishing dissociation amplitude. However, the scale of this effect is much smaller than $J_i$, allowing a broad window, $J_i^{(2)}<\sqrt{\lambda}<J_i$, where the power-law is valid.
\end{thebibliography}
\end{document}